\documentclass[aps,prl,reprint,superscriptaddress,floatfix]{revtex4-1}
\usepackage{graphicx} 
\usepackage{color,dcolumn,bm,amssymb,amsmath} 
\usepackage{soul,xcolor} 
\usepackage{siunitx} 
\usepackage{epstopdf,gensymb} 
\usepackage{hyperref} 
\DeclareSIUnit\angstrom{\text{\AA}} 

\begin{document}

\title{Universal Magnetic Phases in Twisted Bilayer MoTe\textsubscript{2}}

\author{Weijie Li}\thanks{These authors contributed equally}
\affiliation{Department of Physics, University of Washington, Seattle, Washington 98195, USA}
\author{Evgeny Redekop}
\thanks{These authors contributed equally}
\affiliation{Department of Physics, University of California Santa Barbara, Santa Barbara, CA, USA}
\author{Christiano Wang Beach}\thanks{These authors contributed equally}
\affiliation{Department of Physics, University of Washington, Seattle, Washington 98195, USA}
\author{Canxun Zhang}\thanks{These authors contributed equally}
\affiliation{Department of Physics, University of California Santa Barbara, Santa Barbara, CA, USA}
\author{Xiaowei Zhang}
\affiliation{Department of Materials Science and Engineering, University of Washington, Seattle, Washington 98195, USA}
\author{Xiaoyu Liu}
\affiliation{Department of Materials Science and Engineering, University of Washington, Seattle, Washington 98195, USA}
\author{Will Holtzmann}
\affiliation{Department of Physics, University of Washington, Seattle, Washington 98195, USA}
\author{Chaowei Hu}
\affiliation{Department of Physics, University of Washington, Seattle, Washington 98195, USA}
\author{Eric Anderson}
\affiliation{Université Paris-Saclay, CNRS, Centre de Nanosciences et de Nanotechnologies (C2N), 91120 Palaiseau, France}
\author{Heonjoon  Park}
\affiliation{Department of Physics, University of Washington, Seattle, Washington 98195, USA}
\author{Takashi Taniguchi}
\affiliation{Research Center for Materials Nanoarchitectonics, National Institute for Materials Science, 1-1 Namiki, Tsukuba 305-0044, Japan}
\author{Kenji Watanabe}
\affiliation{Research Center for Electronic and Optical Materials, National Institute for Materials Science, 1-1 Namiki, Tsukuba 305-0044, Japan}
\author{Jiun-haw Chu}
\affiliation{Department of Physics, University of Washington, Seattle, Washington 98195, USA}
\author{Liang Fu}
\affiliation{Department of Physics, Massachusetts Institute of Technology, Cambridge, Massachusetts 02139, USA}
\author{Ting Cao}
\affiliation{Department of Materials Science and Engineering, University of Washington, Seattle, Washington 98195, USA}
\author{Di Xiao}
\affiliation{Department of Materials Science and Engineering, University of Washington, Seattle, Washington 98195, USA}
\affiliation{Department of Physics, University of Washington, Seattle, Washington 98195, USA}
\author{Andrea F. Young}
\email{andrea@physics.ucsb.edu}
\affiliation{Department of Physics, University of California Santa Barbara, Santa Barbara, CA, USA}
\author{Xiaodong Xu}
\email{xuxd@uw.edu}
\affiliation{Department of Physics, University of Washington, Seattle, Washington 98195, USA}
\affiliation{Department of Materials Science and Engineering, University of Washington, Seattle, Washington 98195, USA}

\maketitle
\textbf{Twisted bilayer MoTe$_2$ (tMoTe$_2$) has emerged as a robust platform for exploring correlated topological phases, notably supporting fractional Chern insulator (FCI) states at zero magnetic field across a wide range of twist angles. The evolution of magnetism and topology with twist angle remains an open question. Here, we systematically map the magnetic phase diagram of tMoTe$_2$ using local optical spectroscopy and scanning nanoSQUID-on-tip (nSOT) magnetometry. We identify spontaneous ferromagnetism at moir\'e filling factors $\nu = -1$ and $-3$ over a twist angle range from $2.1\degree$ to $3.7\degree$, revealing a universal, twist-angle-insensitive ferromagnetic phase. At $2.1\degree$, we further observe robust ferromagnetism at $\nu = -5$, absent in the devices with larger twist angle --- a signature of the flattening of higher bands in this twist angle range. Temperature-dependent measurements reveal a contrasting twist-angle dependence of the Curie temperatures between $\nu = -1$ and $\nu = -3$, indicating distinct interplay between exchange interaction and bandwidth for the two Chern bands. Despite spontaneous time-reversal symmetry breaking, we find no evidence of a topological gap at $\nu = -3$; however, fragile correlated topological phases could be obscured by the device disorder evident in our spatially resolved measurements. Our results establish a global framework for understanding and controlling magnetic order in tMoTe$_2$ and highlight its potential for accessing correlated topological phases in higher energy Chern band.}

The advent of moir\'e superlattices in two-dimensional (2D) van der Waals materials has opened new avenues for exploring strongly correlated quantum phenomena, including time-reversal symmetry breaking topological phases in the absence of an external magnetic field\cite{cao_correlated_2018, cao_pauli-limit_2021, park_tunable_2021, xie_fractional_2021,sharpe_emergent_2019,serlin_intrinsic_2020, tschirhart_imaging_2021}. 
Twisted bilayers of transition metal dichalcogenides (TMDs) have been established theoretically to be a promising platform due to the combination of strong spin-orbit coupling and flat electronic bands, which can give rise to spontaneous time-reversal symmetry breaking and non-trivial topology\cite{li_spontaneous_2021,wu_topological_2019,zhang_nearly_2019,devakul_magic_2021,yu_giant_2020}. Fractional Chern insulators (FCIs)\cite{spanton_observation_2018,xie_fractional_2021} have been experimentally observed at zero magnetic field in twisted bilayers of MoTe$_2$ (tMoTe$_2$) in a range of twist angles between 2.1$\degree$ and 4$\degree$, establishing this system as a favorable venue for exploring novel strongly-correlated states of matter\cite{cai_signatures_2023,zeng_thermodynamic_2023,park_observation_2023,xu_observation_2023,ji_local_2024,redekop_direct_2024}. Improvements in material quality have led to the observation of deeper sequences of FCI states \cite{park_observation_2025} as well as signatures of superconductivity\cite{xu_signatures_2025}. These discoveries highlight tMoTe$_2$’s potential to emulate the physics of Landau levels in the absence of an external magnetic field.

\begin{figure*}[ht!]
\centering
\includegraphics{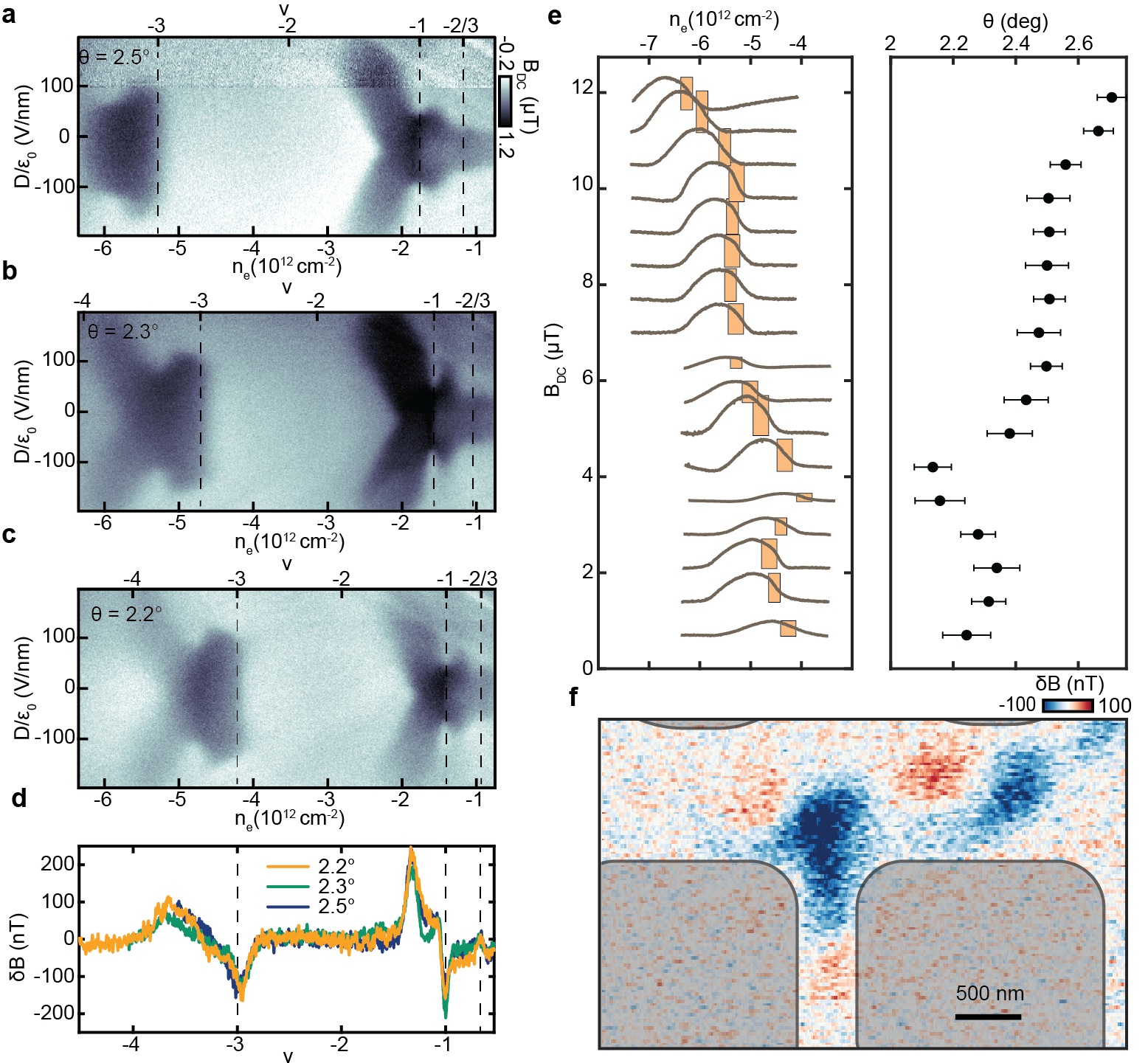}
\caption{\textbf{nanoSQUID on tip microscopy of small angle tMoTe$_2$.} \textbf{a},
Fringe magnetic field $B_{DC}$ as a function of electron density n$_e$ and displacement field $D/\epsilon_0$ over a region of effective twist angle   2.5$\degree$, \textbf{b}, 2.3$\degree$ and \textbf{c}, 2.2$\degree$. Ferromagnetic regions around $\nu = -1$ and $\nu = -3$ persist with varying twist angle and occupy approximately the same area of parameter space, scaled to the area of the moir\'e unit cell. 
\textbf{d}, $\delta B$  along D = 0~V/nm extracted by numerical differentiation of data in panels a-c. 
\textbf{e}, B$_{DC}$ for D = 0~V/nm near $\nu = -3$ at different spatial locations.  The right panel shows the effective local twist angle.  Curves are offset vertically by 0.7~$\mu$T for clarity, and shaded regions indicate range within which $\nu = -3$ lies, given experimental error (see Extended Data Fig. \ref{fig:nSOTLLs} for underlying analysis). \textbf{f}, spatial scan of $\delta B$ at $n_e = -4.5\times 10^{-12} cm^{-2}$ and $D=0$.}
\label{fig:fig2}
\end{figure*}

Despite this progress, the dependence of magnetic properties---including Curie temperature as well as electric field and doping dependence---on the twist angle has not been systematically studied. In particular, the magnetism of the higher energy Chern bands and their dependence on the twist angle are poorly understood\cite{kang_evidence_2024}. Recent theoretical studies have suggested that the bandwidth of all Chern bands at the small twist angle limit decreases with decreasing twist angle \cite{wu_topological_2019, mao_transfer_2024}. This reduced bandwidth may be expected to increase the relative importance of the exchange interactions; however, this effect may be offset by increased carrier-carrier separation at lower twist angles, which weakens the scale of the interactions themselves. The outcome of this interplay of competing effects may significantly affect the phase diagram of the spontaneous ferromagnetism and associated topological states in this system\cite{zhang_polarization-driven_2024,qiu_interaction_2023}.  Experimentally, diverse phenomenology has been reported including competing magnetic ground states in the first Chern band for 3.1-3.3 $\degree$ \cite{chang_evidence_2025, chang_emergent_2025} as well as signatures of a `fractional quantum Spin Hall state' at $\nu = -3$ for a 2.1 $\degree$\cite{kang_evidence_2024} sample.  A systematic study of twist angle dependent magnetism can clarify the nature of the intrinsic phase diagram, as well as provide key input data towards engineering correlated topological phases in the longer term.

\begin{figure*}[ht!]
\centering
\includegraphics{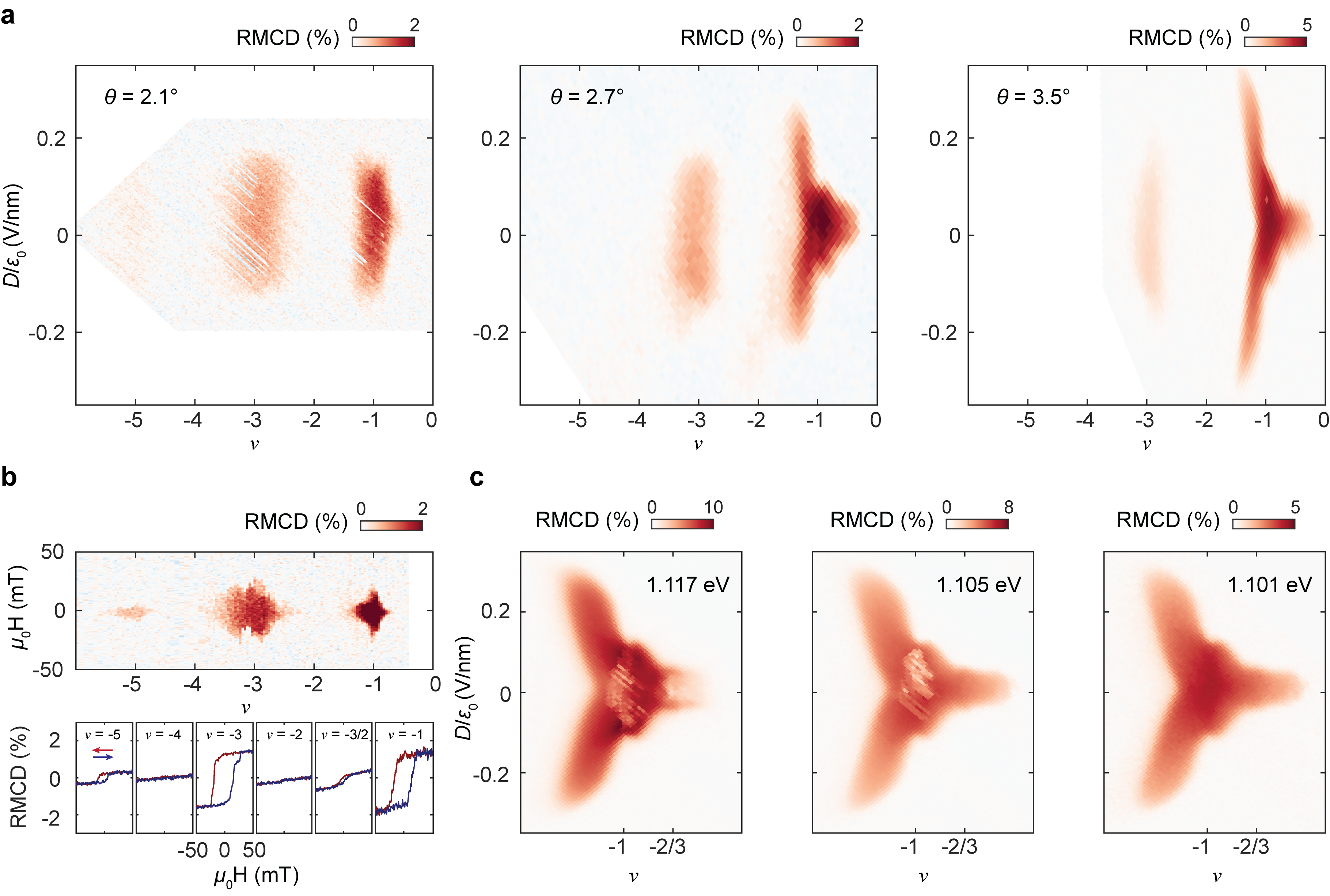} 
\caption{\textbf{Comparison of ferromagnetic phase diagram for different twist angles.} {\bf a.} Reflective magnetic circular dichroism (RMCD) signal as a function of filling factor ($\nu$) and electric field ($D/\varepsilon_0$). The left panel is taken at zero magnetic field ($\mu_0$H) whereas a small magnetic field of 5 mT is applied to suppress magnetic fluctuations for the middle and right panels. RMCD signal can also be seen at $\nu = -5$ for 2.1$\degree$ tMoTe$_2$. Results from additional 2.8$\degree$  and 3.7$\degree$ samples are shown in Extended Data Fig. 4. {\bf b.} Top panel: Hysteretic component of RMCD ($\Delta$RMCD) vs $\nu$ and $\mu_0$H at  2.1$\degree$. $\Delta$RMCD extends beyond the exact integer fillings, spanning across a finite range around $\nu = -1$ and $-3$. Bottom panel: RMCD vs $\mu_0$H at selected $\nu$ as the field is swept down (red) and up (blue). Clear ferromagnetic hysteresis is at $\nu = -1$, $-3/2$, $-3$ and $-5$, while no hysteresis exist at $\nu = -2$ and $-4$. {\bf c.}  RMCD signal near $\nu = -1$ for 3.3$\degree$ tMoTe$_2$ with varied excitation energies of 1.117 eV (left), 1.105 eV (middle) and 1.101 eV (right), respectively.}
\label{fig:fig1}
\end{figure*}

Here, we report that locally, tMoTe$_2$ samples show a seemingly universal magnetic phase diagram across the two lowest energy Chern bands, with distinct magnetic phases in the first ($-1.5\lessapprox \nu\lessapprox -0.4$ ) and second ($-4\lessapprox \nu\lessapprox -2.9$) bands.  
We study hexagonal boron nitride (hBN)-encapsulated tMoTe$_2$ devices with twist angles spanning from 2.1$\degree$ to 3.7$\degree$ (see Methods). Our dual-gated sample geometry allows independent tuning of carrier density ($n_e$) and perpendicular electric field ($D/\epsilon_0$) (see Methods). To characterize the magnetic and topological properties of correlated states, we employ a combination of nanoSQUID-on-tip (nSOT) magnetometry, reflective magnetic circular dichroism (RMCD) and photoluminescence (PL). All measurements are conducted at $T\approx 1.6$~K. In the optical devices, the tMoTe$_2$ bilayer is electrically grounded via a thin graphite layer to reduce the strain induced by electrical contacts and ensure a homogeneous sample area ($\sim 5~\mu m^2$). The nSOT measurement focuses on a separate transport device, used in Ref. \cite{park_ferromagnetism_2025}, with Pt electrodes. Pt contact introduces strain on the MoTe$_2$ flakes which leads to twist angle inhomogeneity. In this sample, the local twist angle varies smoothly from 2.2$\degree$ to 2.8$\degree$, allowing systematic investigation of magnetic behavior as the twist angle gradually varies. \\

\textbf{NanoSQUID-on-tip Microscopy}\\

To clarify the evolution of magnetic order in tMoTe$_2$ across different twist angles we first employ scanning nSOT magnetometry. This technique provides spatial resolution of approximately 100~nm (Methods) with magnetic field sensitivity as good as 0.3~nT/$\sqrt{Hz}$\cite{redekop_direct_2024}. Figures \ref{fig:fig2}a-c show phase diagrams of the out-of-plane fringe magnetic field ($B_{DC}$) acquired at three distinct locations with local twist angles of 2.5$\degree$, 2.3$\degree$, and 2.2$\degree$, respectively. In all three cases, pronounced magnetic signals appear near integer fillings $\nu = -1$ and $\nu = -3$, indicative of ferromagnetism arising from spontaneous polarization of Chern bands. The magnetic phase near $\nu = -1$ closely resembles previous results obtained at $\theta \approx 3.7\degree$\cite{redekop_direct_2024}, exhibiting a sharp drop in $B_{DC}$ exactly at $\nu = -1$, characteristic of edge states from a Chern insulator with negative Chern number. Additionally, a weaker feature appears around $\nu = -2/3$. This feature has been observed in the ferromagnetic (FM) phase diagram for twist angles above 2.6 $\degree$,  due to the formation of the FCI state. Our measurements reveal a similar -2/3 feature in all measured twist angles, suggesting the physics of the lowest Chern band to be identical across angles from $2.2\degree-3.7\degree$ (see Fig. \ref{fig:fig2}d).

\begin{figure*}[ht!]
\centering
\includegraphics{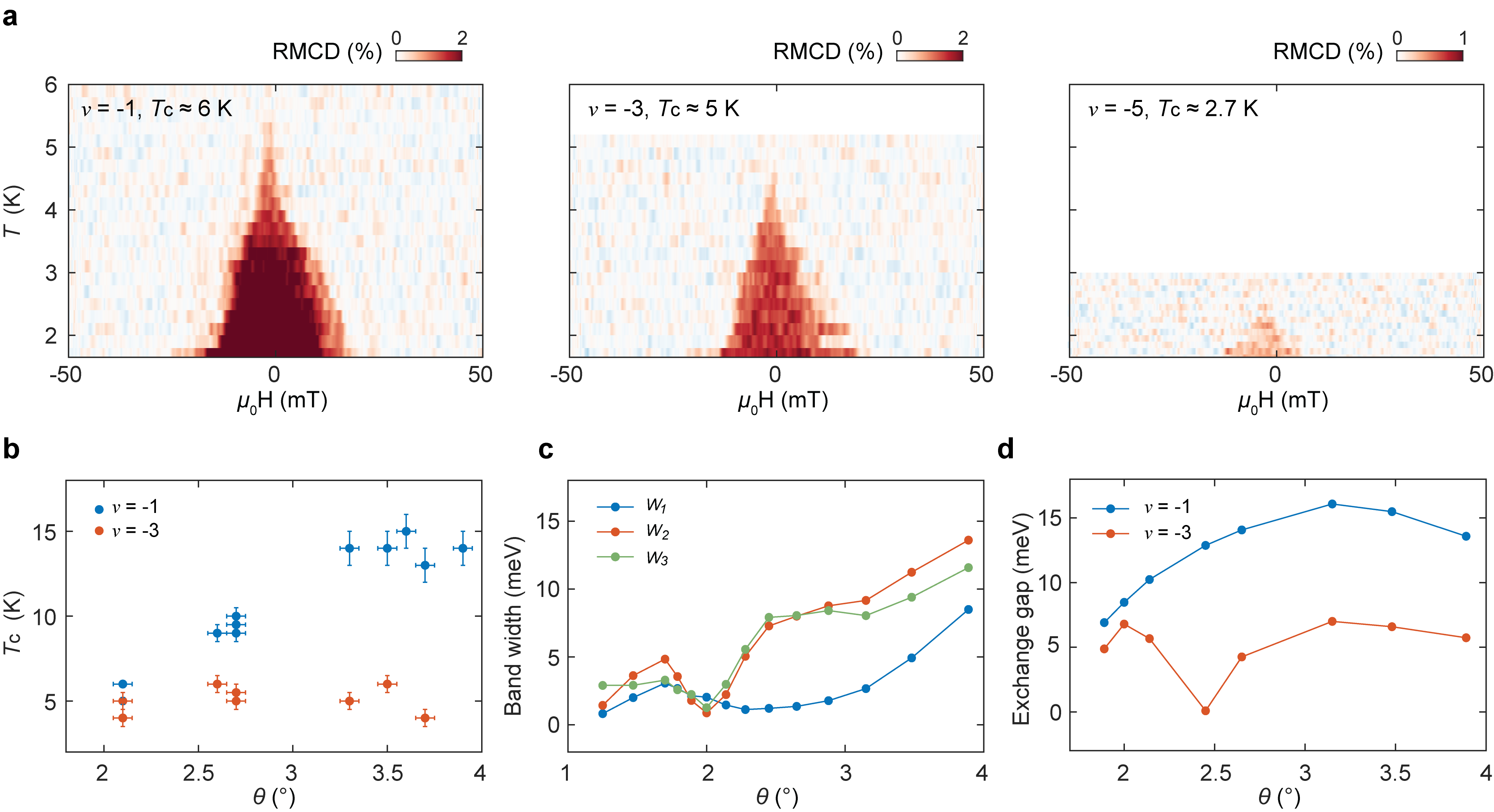} 
\caption{\textbf{Curie temperature at $\nu = -1$ and $-3$.} {\bf a.} $\Delta$RMCD as a function of $\mu_0$H and temperature $T$ at $\nu = -1$ (left), $\nu = -3$ (middle), and $\nu = -5$ (right). Curie temperature $T_c$ at $\nu = -1$, $-3$ and $-5$ are 6 K, 5 K and 2.7 K, respectively. {\bf b.} Measured Curie temperature $T_c$ at $\nu = -1$ (blue) and $\nu = -3$ (red) versus twist angle ($\theta$). $T_c$ at $\nu = -1$ (blue) decreases significantly with smaller twist angle whereas $T_c$ at $\nu = -3$ (red) is relatively stable. The twist angle error bars are $\pm$ 0.05$\degree$(see Methods). The temperature error bars are from the temperature step size. {\bf c.} Twist angle dependence of calculated band width of the first moir\'e band (blue), second moir\'e band (red) and third moir\'e band (green). \textbf{d}, Calculated exchange gap at $\nu = -1$ (blue) and $-3$ (red) versus $\theta$. }
\label{fig:fig3}
\end{figure*}

In contrast, the FM region around $\nu = -3$ lacks pronounced internal features, indicating the absence of topological gaps. The FM phase exhibits a sharp onset that is independent of the displacement field up to $\pm$100mV/nm and is centered at $\nu = -3$. At finite displacement fields, its extent reaches down to $\nu = -4$.
To precisely determine local filling factors, we use the well-defined $\nu = -1$ gap as well as the edge of the single-particle gap determined via chemical potential sensing to reconstruct the filling factor from the applied gate voltages across the experimental doping range (Methods and Extended Data Fig. \ref{fig:nSOTLLs}). 
Figure \ref{fig:fig2}e shows line cuts of $B_{DC}$ versus electron density at zero displacement field ($D = 0$), acquired from various locations across the sample. Shaded regions mark the estimated location of $\nu = -3$. Across all measured locations---spanning twist angles from 2.2$\degree$ to 2.8$\degree$---the edge of the ferromagnetic phase associated with the second moir\'e band coincides with $\nu = -3$, regardless of local variations. In contrast to the broader ferromagnetic pocket around $\nu = -1$, $\nu = -3$ tracks the onset of ferromagnetism, supporting the idea of a robust, twist-angle-insensitive instability towards valley polarization at half-filling of the second moir\'e band.

\begin{figure*}[ht!]
\centering
\includegraphics{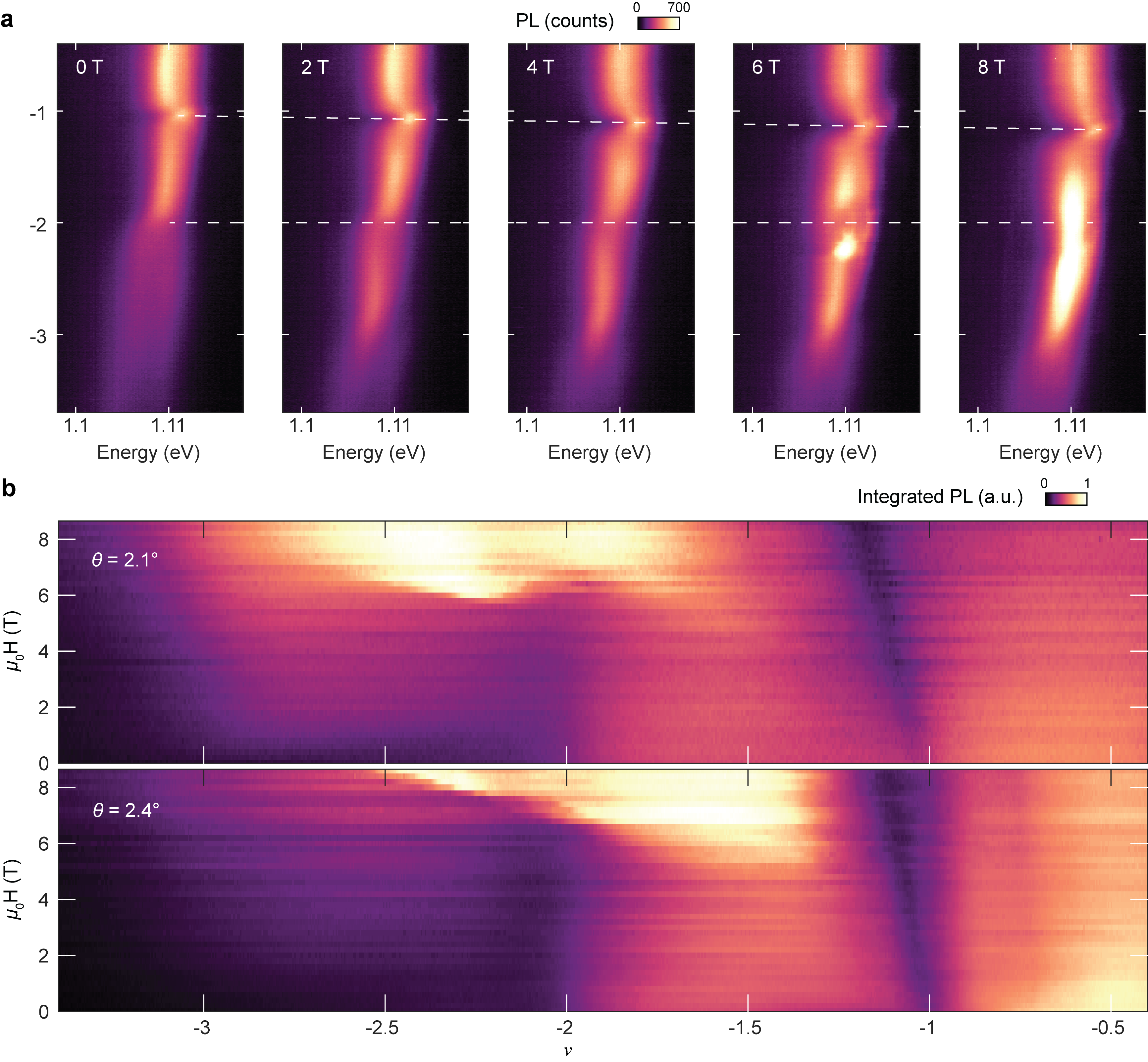} 
\caption{\textbf{Magnetic Field Dependence of the $\nu = -1$ and $-3$ states.} {\bf a.} PL intensity plot versus $\nu$ and photon energy at selected magnetic fields. The dashed lines are guides to the eye. The data is taken from a 2.1$\degree$ sample. {\bf b.} Optically detected fan diagram from spectrally integrated PL intensity versus $\mu_0$H and filling factor $\nu$ in 2.1$\degree$ and 2.4$\degree$ tMoTe$_2$ with 1 meV energy integration window.}
\label{fig:fig4}
\end{figure*}

Recent studies have suggested the emergence of a time-reversal symmetry-breaking Fractional Quantum Spin Hall phase at $\nu = -3$ in the device with 2.1$\degree$ twist angle\cite{kang_evidence_2024,abouelkomsan_non-abelian_2024}. 
Based on our findings, the ferromagnetic phase diagram at $\nu= -3$ , i.e. its shape as a function of $|D/\epsilon_0|$ and $\nu$, is similar across all measured twist angles. The lack of internal structure and the same doping dependence across the whole twist angle range in Fig.\ref{fig:fig2}e, suggests the state at $\nu=-3$ at  2.1$\degree$ is the same as that at larger twist angles. Recent transport measurements at 2.6$\degree$ demonstrate an incipient Chern insulator at $\nu=-3$ \cite{park_ferromagnetism_2025}, i.e. the ground state is gapless at zero magnetic field, with a Chern insulator stabilized at high magnetic field. We note that because ferromagnetism onsets so close to $\nu=-3$, transport phenomenology at this filling will be particularly susceptible to twist angle disorder. This is evident in the spatial scan of $\delta B\propto \delta B/\delta n_e$, shown in Fig. \ref{fig:fig2}f. Here, blue regions are in the vicinity of the transition region near $\nu=-3$, while red and white regions are significantly detuned. Transport experiments performed between contacts (shown in gray) may thus sample this microscopically inhomogeneous physics in unpredictable fashion.  
Improved material quality, particularly reduction in strain, is thus likely to play a key role in definitively resolving the nature of the B=0 ground state at $\nu = -3$.\\

\textbf{Robust Magnetic Phases in tMoTe$_2$}
\\

To further probe the magnetic phase diagram across devices with a larger twist angle range, we measure RMCD as a function of filling factor $\nu$ and electric field $D/\epsilon_0$. As shown in Figure \ref{fig:fig1}a, non-zero RMCD signal around $\nu = -1$ and $-3$ is observed for 2.1$\degree$ (left), 2.7$\degree$ (middle) and 3.5$\degree$ (right) tMoTe$_2$, indicating the presence of ferromagnetism at these integer fillings across a broad twist angle range. Furthermore, at $\nu = -5$, a pronounced RMCD signal is detected only in the device with 2.1$\degree$ twist angle but not with 2.7$\degree$. This is consistent with the theoretical predictions that the moir\'e bandwidth reduces with lowering the twist angle. For smaller angles, higher bands are sufficiently flat and isolated to support a sequence of at least three consecutive Chern bands\cite{wang_higher_2024,zhang_non-abelian_2024,zhang_polarization-driven_2024}.

Ferromagnetism at $\nu = -1$, $-3$ and $-5$ in 2.1$\degree$ tMoTe$_2$ is further characterized by magnetic hysteresis measurements in Figures \ref{fig:fig1}b, where forward and backward magnetic field sweeps of RMCD exhibit a clear hysteresis loop --- a hallmark of spontaneously broken time reversal symmetry. Additional electric field dependent data are shown in Extended Data Fig. \ref{fig:another2p1}. Notably, the ferromagnetism extends beyond exact integer fillings, spanning over a finite range above $\nu = -1$ and $-3$. This is necessary for hosting FCI states in higher energy Chern bands. However, no gapped states at fractional filling are observed in our experiment. If these states exist, it is possible that the base temperature (1.6 K) of the experiment or material disorder prevents their observation. At $\nu = -2$ and $\nu = -4$, the RMCD signal vanishes, consistent with a scenario in which the conjugate Chern bands in opposite valleys are fully filled, resulting in the disappearance of the total spin-valley polarization and the RMCD signal. 

We note that when utilizing RMCD as a probe of ferromagnetism in tMoTe$_2$, it is critical to minimize perturbations arising from resonant optical excitation. Figure \ref{fig:fig1}c illustrates such an excitation energy dependent effect for a 3.3$\degree$ sample. The RCMD phase diagrams are shown with 1.117 eV (above resonance), 1.105 eV (on resonance), and 1.101 eV (below resonance) with the same excitation power. For the slightly above resonant excitation, the RMCD signal is suppressed near -1, and -2/3 and below for small electric fields, and revives near the electric field induced phase transition boundary\cite{anderson_programming_2023,cai_signatures_2023,park_observation_2023,xu_observation_2023,zeng_thermodynamic_2023,redekop_direct_2024,ji_local_2024,anderson_trion_2024}. This behavior is similar to the recent report in devices with similar twist angles, which is attributed to anomalous AFM states\cite{chang_emergent_2025, chang_evidence_2025}. For the on-resonance excitation, RMCD is only suppressed for a small phase space around $\nu = -1$ state.  Lastly, RMCD behaves normally for the excitation slightly below the resonance. The observed dependence of RMCD on excitation energy is likely the result of optically induced domain flipping due to circular polarized optical pumping of valley polarization. To avoid such a perturbative effect, RMCD measurements across different twist angles were all conducted with excitation energies below trion resonance.\\

\textbf{Critical Temperature of Ferromagnetism}
\\

Having established the existence of spontaneous ferromagnetism at $\nu = -1$ and $-3$, we next examine its twist angle-dependent Curie temperature. Figure \ref{fig:fig3}a shows the temperature dependence of the hysteretic component ($\Delta$RMCD) at $\nu = -1$ (left), $\nu = -3$ (middle) and $\nu = -5$ (right) in the 2.1$\degree$ tMoTe$_2$ device. The critical temperature $T_c$ is defined as the temperature at which $\Delta$RMCD vanishes, corresponding to the Curie temperature of the underlying ferromagnetic phase. We find $T_c\approx$ 6~K at $\nu = -1$ and 5~K at $\nu = -3$ (see data from another 2.1$\degree$ device in Extended Data Fig. \ref{fig:rmcdtemp}), whereas the ferromagnetism at $\nu = -5$ exhibits a much lower $T_c\approx$ 2.7~K. 
The highest $T_c$ observed at $\nu = -1$ is expected due to the lower screening effect at lower doping, and consistent with previous reports at larger twist angles, attributed to stronger magnetic exchange interaction at $\nu = -1$ \cite{cai_signatures_2023,zeng_thermodynamic_2023,park_observation_2023,xu_observation_2023,ji_local_2024,redekop_direct_2024,anderson_programming_2023}.

A systematic study of twist-angle dependence of $T_c$ at $\nu = -1$ and $-3$ is summarized in Fig. \ref{fig:fig3}b. At $\nu = -1$, $T_c$ increases from 5-6~K at 2.1$\degree$ to $\sim 10$ K near 2.6$\degree$ and eventually saturates at 14~K. 
This trend is consistent with our Hartree-Fock calculations of the exchange gap (see Methods), shown in Fig.~\ref{fig:fig3}d.
Note that the bandwidth of the first moiré band shows an overall increase with twist angle (Fig.~\ref{fig:fig3}c), but the reduced moiré wavelength decreases the carrier separation and leads to a higher $T_c$. 
In contrast, $T_c$ at $\nu = -3$ remains relatively constant at 4–6~K across the twist angle range from 2.1$^\circ$ to 3.7$^\circ$. This weak angle dependence is consistent with the nearly identical magnetic response observed at $\nu = -3$, as discussed previously.
Our Hartree-Fock calculations show that the exchange gaps at $\nu = -1$ and $-3$ are comparable at small twist angles, but at larger angles, the gap at $\nu = -1$ becomes significantly larger---mirroring the observed behavior of $T_c$ in both regimes. The calculations also reveal a dip in the exchange gap near 2.5$^\circ$, which originates from a band inversion between the second and third moiré bands. Experimentally, pinpointing this critical angle is challenging, but outside this region, the exchange gap at $\nu = -3$ exhibits much weaker twist-angle dependence than that at $\nu = -1$. \\

\textbf{Magnetic Field Dependence of Correlated States}
\\

To probe the topology of the moir\'e bands at 2.1$\degree$, we perform the filling-dependent trion PL spectroscopy at effective zero displacement field and as a function of applied magnetic field $\mu_0$H (Figure \ref{fig:fig4}a ). At $\nu = -1$ and $\mu_0$H = 0 T, we observe a suppression in PL intensity and a blueshift of the spectral peak. As magnetic field increases, the density corresponding to this feature shifts, consistent with prior reports in 3.7$\degree$ tMoTe$_2$\cite{cai_signatures_2023}. At $\nu = -2$, PL emission is also significantly suppressed. An additional blue peak appears at filling factors above $\nu = -2$, where the first moir\'e band is fully occupied, and holes populate the second moir\'e band. We attribute this feature to trions formed by excitons localized at MM sites, given that MX and XM sites are already filled by holes. This additional feature is also observed at 2.7$\degree$ and 3.3$\degree$ samples (Extended Data Fig. \ref{fig:PLother}). At finite magnetic fields, PL spectra exhibit enhanced emission between $\nu=-2$ and $-3$. For instance, a drastically enhanced PL is observed just above $\nu = -2$ for $\mu_0$H$> 6 T$.  There is also a smooth intensity drop for $\nu$ near and above -3, which is particularly appreciable as magnetic field increases. 

We construct a fan diagram by spectrally integrating PL intensity over the peak and plotting the results versus $\nu$ and $\mu_0$H (Figure \ref{fig:fig4}b, top panel). The linear dependence of the carrier density of the intensity dip on the magnetic field at $\nu = -1$ matches the Streda slope of $C = -1$, in agreement with observations at larger twist angles\cite{cai_signatures_2023,zeng_thermodynamic_2023,park_observation_2023,xu_observation_2023}. At $\nu = -2$, the dip is a non-dispersive feature ($C = 0$) at low field, and transitions into a bright dispersive feature at high field (above 6T). We attribute the bright feature to a magnetic field induced $C = -2$ Chern insulator state, as reported in 2.6$\degree$ tMoTe$_2$\cite{park_ferromagnetism_2025} (see Extended Data Figure. \ref{fig:opticalfanderivative}). At $\nu=-3$, although there is no dip as at $\nu=-1$, we observe the boundary between the bright and dark PL shifts versus $\mu_0$H. This may suggest the state is an incipient Chern insulator state, as observed in larger twist angle \cite{park_ferromagnetism_2025}.  We also obtained optically detected fan diagrams of  2.4$\degree$ (Figure 4b, bottom panel) and 2.8$\degree$ devices (Extended Data Figure \ref{fig:opticalfanother}). The results are similar to the 2.1$\degree$ device, which reinforces our speculation that the $\nu = -3$ at 2.1$\degree$ shares the same topological origin as larger twist angles. Further transport measurements versus magnetic field at 2.1$\degree$ is necessary to confirm our interpretation.

\textbf{Conclusions}


Our study confirms the presence of the phases with spontaneously broken time-reversal symmetry around $\nu = -1$ and $\nu=-3$ in tMoTe$_2$ across a wide range of twist angles (2.1-3.7$\degree$). The Curie temperature at these two fillings have distinct dependence on the twist angle, stemming from the competition between bandwidth and direct exchange interactions. We further provide compelling evidence that the higher Chern bands become increasingly flat as the twist angle decreases. However, we do not observe signatures of a topological gap at $\nu=-3$ in either of the techniques used in this study, which implies the trivial nature of the state at $\nu=-3$ with zero applied magnetic field. We expect improved device and material quality to be crucial in observing topological phases of matter in higher Chern bands of tMoTe$_2$.

\bibliographystyle{custom}
\bibliography{references}

\section*{Methods}

\subsection*{Device fabrication}
The optical devices were fabricated using graphite, hBN, and MoTe$_2$, which were mechanically exfoliated onto Si/SiO$_2$ substrates and identified via optical microscopy. The hBN thickness was measured using atomic force microscopy to determine carrier density and electric field. Device assembly began with the preparation of the back gate by sequentially picking up a bottom hBN dielectric, a bottom gate graphite electrode and then dropping down onto the substrate at ~170 $\degree$C using the PC dry-transfer techniques. The heterostructure was then immersed in chloroform for 10 minutes to dissolve the polymer, followed by AFM cleaning to remove residual PC. Next, a thin hBN flake, a strip of contact graphite, and half of a monolayer MoTe$_2$ flake were sequentially picked up. The second half of the MoTe$_2$ monolayer was then rotated to the target twist angle before being incorporated, forming the moir\'e heterostructure, which was subsequently melted onto the prepared back gate. As MoTe$_2$ is air sensitive, its exfoliation and the tMoTe$_2$ assembly were both carried out in an argon filled glovebox. To ensure a clean interface, we use AFM cleaning to squeeze out trapped gas bubbles, which is essential to achieve a homogenous sample region and optimal optical response, particularly for small-twist-angle devices. Finally, top gate graphite and top hBN dielectric are transferred on top of the whole stack before patterning gold wire bonding pads using standard electron beam lithography, followed by E-beam evaporation to complete the device fabrication.

\subsection*{Optical measurements}
\subsubsection*{RMCD and photoluminescence measurements}
All measurements were performed in a closed-loop magneto-optical cryostat (attoDRY 2100XL) with an attocube xyz piezo stage and xy scanners, a 9T z-axis superconducting magnet, and with a base temperature of 1.65K. 

RMCD measurements were taken in resonance with the trion feature by filtering a broadband supercontinuum source (NKT SuperK Fianium FIR-20) through a homebuilt filter to achieve a narrow excitation bandwidth ($\sim$1 nm). The out-of-plane magnetization of the sample induces a MCD signal $\Delta$R, the difference between the reflected right and left-circularly polarized light. To obtain the normalized RMCD $\Delta$R/R, the laser intensity was chopped at p = 950Hz and the phase was modulated by $\lambda$/4 via a photoelastic modulator at f = 50kHz. An InGaAs avalanche photodiode detector was used to collect the reflected signal, and the output was read by two lock-in amplifiers (SR830 and SR860). The ratio between the p-component signal I$_p$ and f-component signal I$_f$ gives the RMCD signal: $\Delta$R/R = I$_f$ /(J$_1$($\pi$/2) × I$_p$) where J$_1$ is the first-order Bessel function.

The photoluminescence measurements were taken with linearly polarized 632.8 nm HeNe laser excitation focused on the sample by a high-NA nonmagnetic cryogenic objective to be a $\sim$1 $\mu$m beam spot. Sample PL emission was collected by the same objective and passed to the spectrometer, where the PL signal is dispersed with a diffraction grating (Princeton Instruments, 300 grooves/mm at 1.2 $\mu$m blaze) and detected by a liquid nitrogen cooled infrared InGaAs CCD (Princeton Instruments PyLoN-IR 1.7).

\subsubsection*{Determination of doping density and electric field}
In optical measurements, we use a parallel plate capacitor model to determine the carrier density n$_e$ and displacement field $D$/$\varepsilon_0$ from the applied top and bottom gate voltages, V$_{tg}$ and V$_{bg}$. Gate capacitances C$_{tg}$ and C$_{bg}$ are obtained using the hBN thickness determined by atomic force microscopy measurements, taking the hBN dielectric constant to be 3.0. While we only consider bottom hBN dielectric thickness for C$_{bg}$, we count both the top hBN dielectric and the thin hBN for C$_{tg}$. Thus, $n$ and $D$ can be computed as n$_e$ = (V$_{tg}$C$_{tg}$ + V$_{bg}$C$_{bg}$)/$e$-n$_\text{offset}$ and $D$/$\varepsilon_0$ = (V$_{tg}$C$_{tg}$ - V$_{bg}$C$_{bg}$)/2$\varepsilon_0$, where $e$ is the electron charge and $\varepsilon_0$ is the vacuum permittivity. The carrier density offset n$_\text{offset}$ is derived from fitting to the integer (and fractional) states in PL spectra, further confirmed by examining the PL integrated intensity versus n$_e$ and $D$/$\varepsilon_0$ as well as RMCD hysteresis. The filling factor is subsequently defined by -n$_e$/n$_e (\nu = -1)$ and this assignment is extended to higher fillings. The capacitor model determination of n$_e$ and assigned filling factors can be further validated by tracing the optical Landau fan down to zero magnetic field. The twist angle is estimated from n$_e (\nu = -1)$, which is consistent with the target angle during device fabrication. The uncertainty in the twist angles estimation primarily arises from the uncertainty in determining exact hole densities at correlated states. This leads to an uncertainty in n$_e (\nu = -1)$ of $\sim$5×10$^{10}$ cm$^{-2}$, corresponding to a twist angle error bar of ±0.05$\degree$. There are two 2.1$\degree$ devices D1 and D2 in the main text, with D1 data in Figure 2-3, and D2 data in Figure 4. The hole density at a filling of $\nu = -1$ is 1.24-1.30×10$^{12}$ cm$^{-2}$ for D1 and 1.20-1.25×10$^{12}$ cm$^{-2}$ for D2, that is, the twist angle is 2.1$\degree$±0.05$\degree$ for these two devices.
\subsection*{NanoSQUID measurements}

\subsubsection*{nSOT sensor fabrication and local magnetometry measurements}
We performed local magnetometry using a superconducting quantum interference device on the sharp apex of a quartz pipette (nanoSQUID-on-tip, nSOT). The probe was fabricated following previous protocols[1–3]. A quartz micropipette with a 0.5 mm inner diameter was pulled into a sharp tip with an apex diameter of approximately 150 nm. Titanium/gold coarse contacts (5 nm / 50 nm) were deposited via electron-beam evaporation at a rate of $\SI{2}{\angstrom/s}$. A shunt resistor (8 nm Ti / 15 nm Au) was then deposited within 500 $\mu$m of the tip apex, resulting in a $3~\Omega$ resistance.

The contact pads were coated with a thick layer of indium solder to reduce low-temperature resistance and improve electrical connection to the leaf springs in the tip holder. Finally, indium (T$_c$ = 3.4 K) was deposited in a home-built thermal evaporator from three angles to cover two opposing contacts at 110$\degree$ relative to the apex, followed by head-on deposition. During this process, the tip holder was mounted on a cryostat and shielded by liquid nitrogen jacketing to maintain a temperature below 10 K. Each deposition step was preceded by 5–15 minutes of thermalization with helium exchange gas at a pressure of $5\times 10^{-3}$ mbar. Typical indium film thicknesses were 30–60 nm for side depositions and 20–50 nm for the head-on direction, with a deposition rate of 0.1 nm/s, yielding uniform, fine-grain superconducting films near the apex.

nSOT measurements were carried out in a helium cryostat at a base temperature of 1.7 K. The magnetic flux through the tip was measured using a quasi-voltage-biased readout with a series SQUID array amplifier (SSAA) and compensation circuitry [4]. To calibrate the sensor sensitivity, we measured the voltage noise spectrum of the SSAA and converted it to magnetic noise via the magnetic field-to-voltage transfer function, obtained from the tip’s voltage response to a $20~\mu T$ magnetic field step. The resulting magnetic sensitivity was approximately $1-10~nT/\sqrt{Hz}$, with an effective sensor diameter of ~200 nm, corresponding to $20-200~n\Phi_0/\sqrt{Hz}$.

\subsubsection*{Determination of local filling factor from nSOT measurements}
We used scanning nSOT magnetometry to independently calibrate the local charge carrier density and determine absolute filling factors across the sample. The total carrier density (n$_e$ + n$_\text{offset}$) was calculated using a parallel-plate capacitor model, with gate capacitance determined from the dielectric thickness of the hBN layers as described in the previous session.

To determine the density offset (n$_\text{offset}$) corresponding to charge neutrality ($\nu = 0$), we measured the modulated magnetic response of Landau levels in the top graphite gate. These serve as sensitive chemical potential sensors (Extended Data Fig. \ref{fig:nSOTLLs}a). With the sample grounded, we modulated the displacement field D and measured the resulting fringe field $\delta$B$_D$. A kink in the top-gate Landau fan pattern marked the transition between transparent (incompressible) and screening (hole-doped) regimes of the tMoTe$_2$, providing an accurate estimate of n$_\text{offset}$.

To anchor the filling factor axis, we then measured the ac magnetic response $\delta$B$_n$ to small modulations of n$_e$ at D = 0. A sharp dip in $\delta$B$_n$ marked the Chern insulating gap at $\nu = –1$, attributed to edge states of a Chern insulator with negative Chern number (Extended Data Fig. \ref{fig:nSOTLLs}b). Using this as a reference, we extrapolated the filling factor via:

$\nu$ = –n$_e$ / n$_e$($\nu = –1$)

This procedure enables precise local assignment of filling factor, facilitating consistent interpretation of the spatially resolved magnetic phase diagrams.

\subsubsection*{Measurement conditions for presented data}
nSOT data were acquired in two modes. In the quasi-dc mode, gate voltages were modulated as square waves between a measurement point and a reference point. The resulting fringe field was proportional to B$_\text{DC}$. In the ac mode, small sinusoidal modulations were applied atop static voltages, and the signal was proportional to $\partial$B$_\text{DC}$/$\partial$(V), depending on the modulation direction.

\begin{itemize}
    \item Figs. \ref{fig:fig2}a--d
        \subitem height = 150~nm; \\
        Square wave modulation; \\
        frequency = 201.777~Hz; \\
        Reference: $V_t = -2.4$~V, $V_b = -4.0$~V
    \item Fig. \ref{fig:fig2}e and Extended Data Fig. \ref{fig:nSOTLLs}b (blue curve)
        \subitem height = 100~nm; \\
        Square wave modulation; \\
        frequency = 151.777~Hz; \\
        Reference: $V_t = -2.6$~V, $V_b = -3.67$~V (low twist); $V_t = -3.0$~V, $V_b = -4.33$~V (high twist)
    \item Extended Data Figs. \ref{fig:nSOTLLs}a and \ref{fig:nSOTLLs}b inset
        \subitem height = 100~nm; \\
        Sinusoidal modulation; \\
        frequency = 201.777~Hz; \\
        $\delta V_t = 15.5$~mV, $\delta V_b = 25.8$~mV; \\
        phase difference = 180$\degree$; \\
        $\delta D = 2.5$~mV/nm
    \item Fig. \ref{fig:fig2}f and Extended Data Fig. \ref{fig:nSOTLLs}b (red curve)
        \subitem height = 100~nm; \\
        Sinusoidal modulation; \\
        frequency = 201.777~Hz; \\
        $\delta V_t = 15.5$~mV, $\delta V_b = 25.8$~mV; \\
        phase difference = 0$\degree$; \\
        $\delta n_e \approx \SI{2.8e10}{cm^{-2}}$
\end{itemize}

\subsection*{Numerical methods}
For the band width calculation, the single-particle moiré band structures are computed using density functional theory (DFT) simulations with the SIESTA package\cite{soler_siesta_2002}. Calculations employ optimized norm-conserving Vanderbilt pseudopotentials\cite{hamann_optimized_2013}, the Perdew-Burke-Ernzerhof exchange-correlation functional\cite{perdew_generalized_1996}, and a double-zeta plus polarization basis set. Moiré superlattice relaxations are carried out using machine learning force fields (MLFFs), which are trained via the deep potential molecular dynamics (DPMD) approach\cite{zhang_deep_2018,wang_deepmd_2018}, with training data generated from ab initio molecular dynamics (AIMD) simulations using the VASP package\cite{kresse_efficiency_1996}. Further details on band structure calculations and MLFF parameterization are provided in Ref.\cite{zhang_polarization-driven_2024}.

For the exchange gap calculation at each twist angle, we build a 12-orbital Wannier Hamiltonian—six orbitals per valley—derived from DFT calculations following the method in Ref. \cite{wang_higher_2024}. Using these models, we carry out self-consistent Hartree–Fock calculations at fillings $\nu = -1$ and $\nu = -3$. To temper the Hartree–Fock tendency to overestimate interaction effects, the dual-gate Coulomb potential is screened with a relative dielectric constant $\epsilon$ = 40. For $\nu = -1$ and $\nu = -3$, the resulting gap separates states with opposite spin and is therefore termed the exchange gap.

\textbf{Acknowledgements:} The work at UW is mainly supported by the U.S. Department of Energy (DOE), Office of Science, Basic Energy Sciences (BES), under the award DE-SC0012509. Device fabrication and measurements are partially supported by Vannevar Bush Faculty Fellowship (Award number N000142512047). Bulk MoTe\textsubscript{2} crystal growth and characterization is supported by Programmable Quantum Materials, an Energy Frontier Research Center funded by DOE BES under award DE-SC0019443. T.C. acknowledges the support by the U.S. Department of Energy, Office of Basic Energy Sciences, under Contract No. DE-SC0025327 for part of the theoretical analysis. This research used resources of the National Energy Research Scientific Computing Center, a DOE Office of Science User Facility supported by the Office of Science of the U.S. Department of Energy under Contract No. DE-AC02-05CH11231 using NERSC award BES-ERCAP0032546, BES-ERCAP0033256, and BES-ERCAP0033507.The authors also acknowledge the use of the facilities and instrumentation supported by NSF MRSEC DMR-2308979. K.W. and T.T. acknowledge support from the JSPS KAKENHI (Grant Numbers 21H05233 and 23H02052) , the CREST (JPMJCR24A5), JST and World Premier International Research Center Initiative (WPI), MEXT, Japan. XX acknowledges support from the State of Washington funded Clean Energy Institute and from the Boeing Distinguished Professorship in Physics. 
Work at UCSB was primarily supported by the Army Research Office under award W911NF-20-2-0166. A.F.Y. acknowledges additional support by the Gordon and Betty Moore Foundation EPIQS program under award GBMF9471. 

\textbf{Author contributions: }X.X. and A.F.Y. conceived and supervised the experiment. W.L. fabricated optical samples. W.L. and C.W.B. performed magneto-optical measurements assisted by W.H.. E.A. and H.P. fabricated transport samples for nanoSQUID measurements. E.R. and C.Z. performed the nanoSQUID measurements. C.H., and J.-H.C. synthesized and characterized the bulk MoTe$_2$ crystals. X.Z., T.C., and D.X. performed large-scale DFT calculations. X.L., T.C., and D.X. performed Hartree-Fock calculations. W.L., E.R., C.W.B., C.Z., A.F.Y., and X.X. analyzed and interpreted the results with the inputs from L.F., T.C., and D.X.. T.T. and K.W. synthesized the hBN crystals. X.X., A.F.Y., W.L., E.R., C.W.B., and C.Z. wrote the paper with input from all authors. All authors discussed the results.

\textbf{Competing interests: }The authors declare no competing interests.

\textbf{Data availability: }Source data that reproduces the plots are provided with this paper. All supporting data for this paper and other findings of this study are available from the corresponding authors upon reasonable request.

\clearpage
\setcounter{figure}{0}
\renewcommand{\figurename}{\textbf{Extended Data Fig.}}
\begin{figure*}[ht!]
\centering
\includegraphics{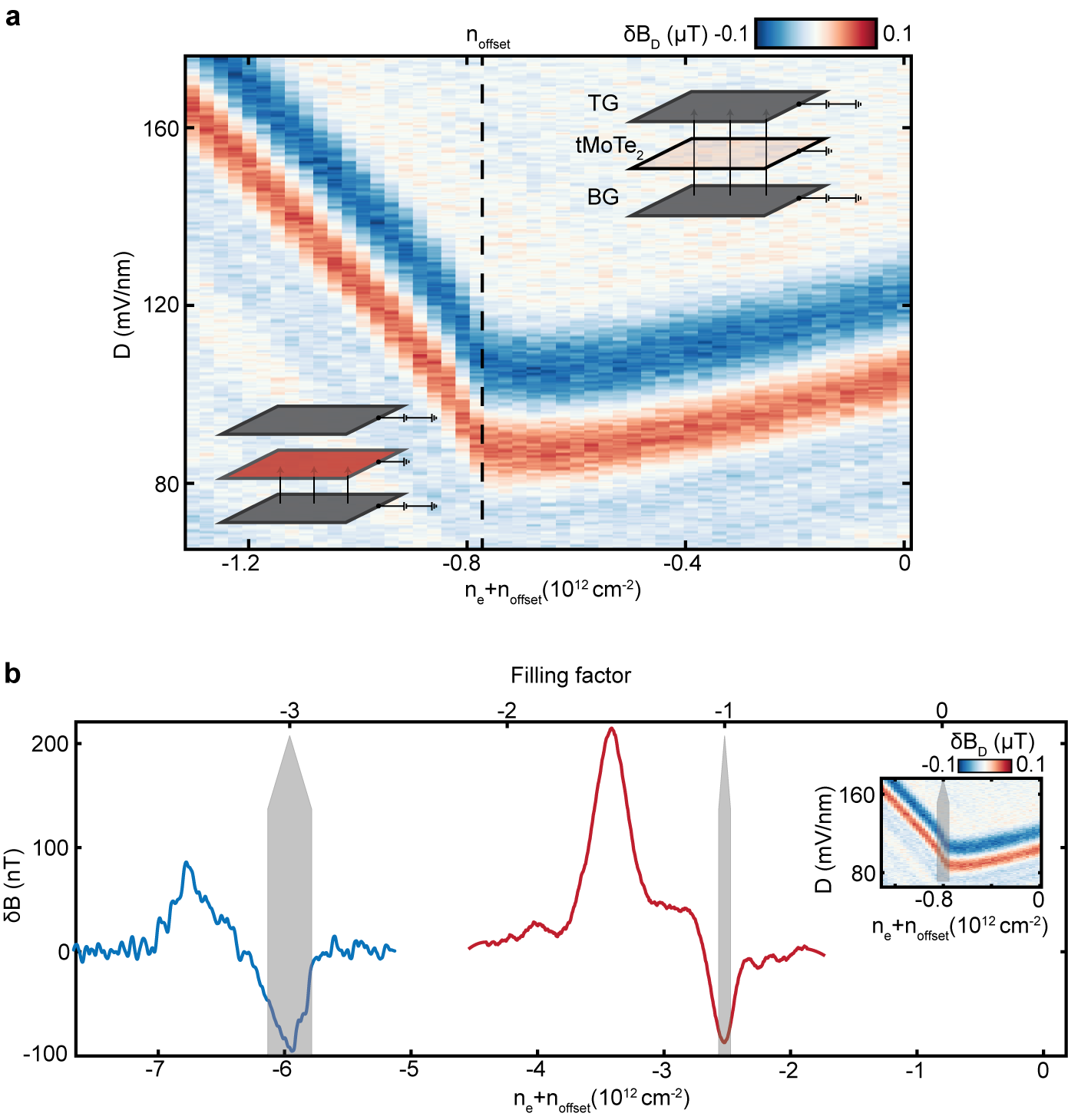}
\caption{\textbf{Local Filling Factor Calibration.} {\bf a.} Modulated magnetic response of Landau levels in the top graphite gate. {\bf b.} AC magnetic response $\delta$B$_n$ to small modulations of n$_e$ at $D/\varepsilon_0$ = 0, with a sharp dip marking the Chern insulating gap at $\nu = –1$. Inset: Additional calibration details.}
\label{fig:nSOTLLs}
\end{figure*}


\begin{figure*}[ht!]
\centering
\includegraphics{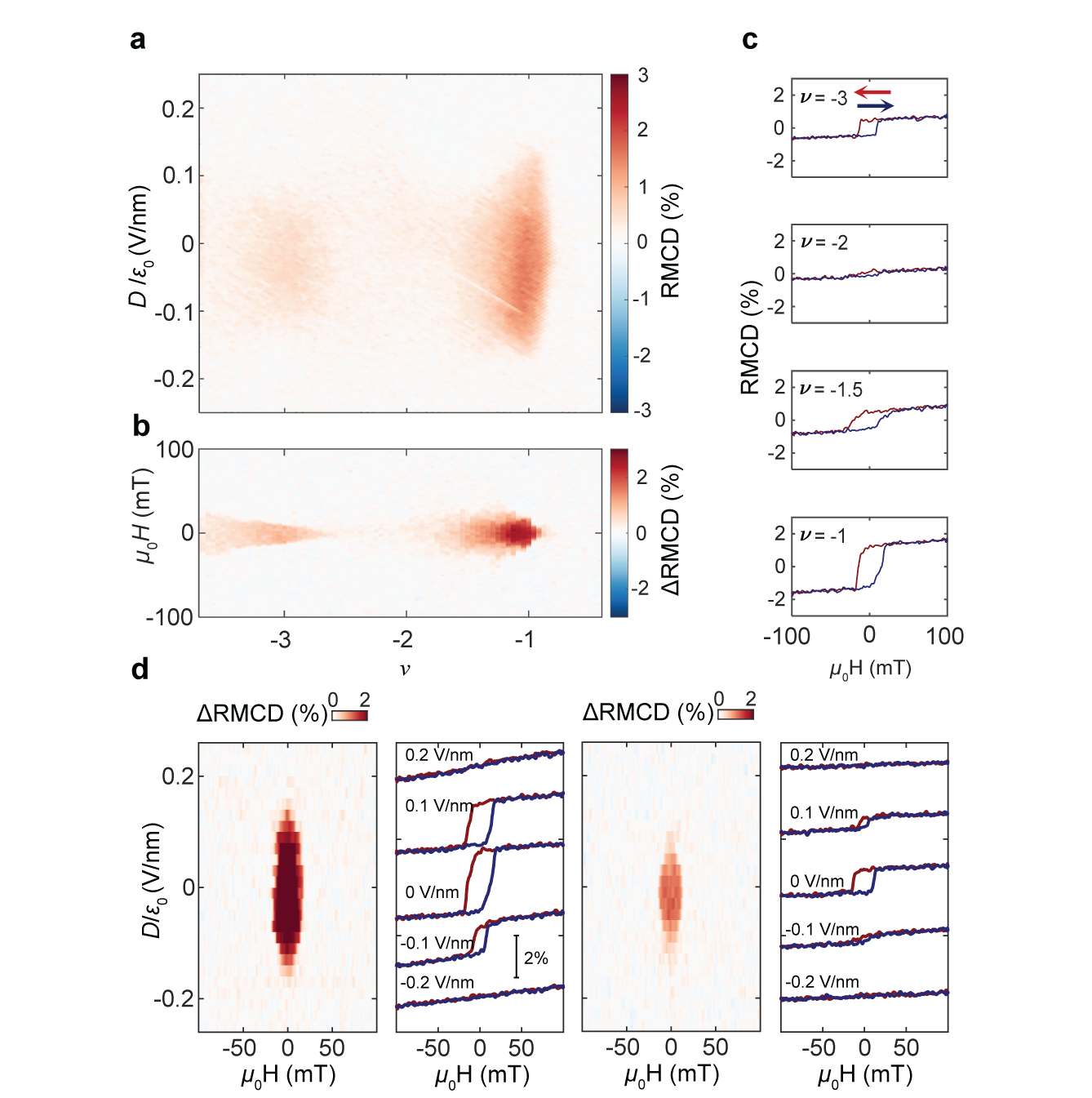}
\caption{\textbf{Electrically tunable spontaneous ferromagnetism at $\nu = -1$ and $-3$ in another 2.1$\degree$ device.} {\bf a.} Reflective magnetic circular dichroism (RMCD) signal as a function of filling factor ($\nu$) and electric field ($D$/$\epsilon_0$). A small magnetic field ($\mu_0$H) of 5 mT is applied to suppress magnetic fluctuations. {\bf b.} Hysteretic component of RMCD ($\Delta$RMCD) vs $\nu$ and $\mu_0$H. $\Delta$RMCD extends beyond the exact integer fillings, spanning from $\nu = -1$ over $-3/2$ and across a finite range around $\nu = -3$. {\bf c.} RMCD vs $\mu_0$H at selected $\nu$ as the field is swept down (red) and up (blue). Clear signatures of ferromagnetism are visible at $\nu = -1$, $\nu = -1.5$ and $\nu = -3$, while it is suppressed at $\nu = -2$. {\bf d.} $\Delta$RMCD as a function of $\mu_0$H and D/$\varepsilon_0$ at $\nu = -1$ (left panel) and $\nu = -3$ (right panel), respectively, highlighting the electric field tunable ferromagnetism. The line-cut plots are $\Delta$RMCD vs $\mu_0$H at selected $D_\text{eff}$/$\varepsilon_0$ = -0.2, -0.1, 0, 0.1, 0.2 V/nm from bottom to top offset for clarity.}
\label{fig:another2p1}
\end{figure*}

\begin{figure*}[ht!]
\centering
\includegraphics{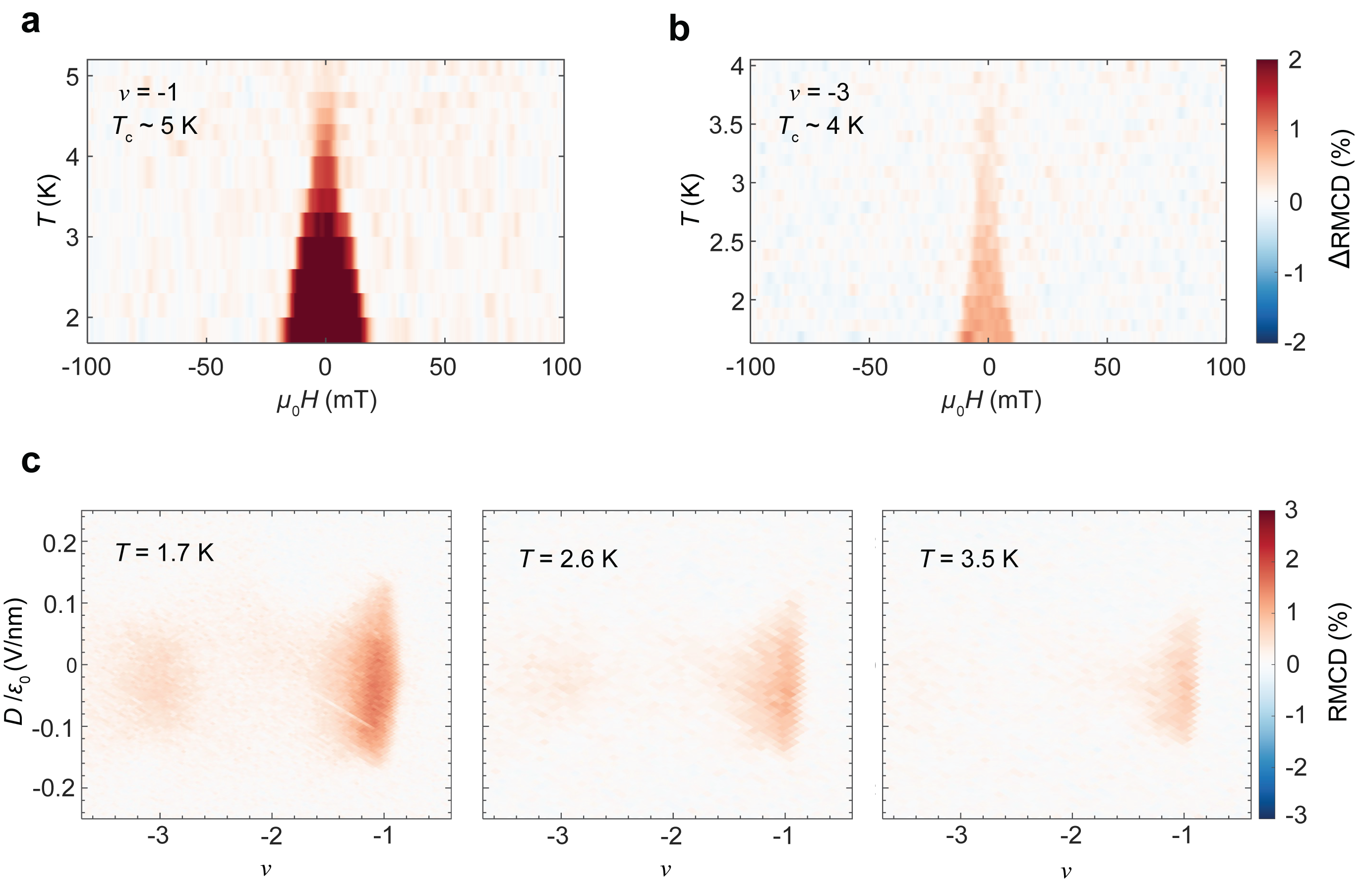}
\caption{\textbf{Critical temperature of ferromagnetism at $\nu = -1$ and $-3$ in another 2.1$\degree$ device.} {\bf a, b.} $\Delta$RMCD as a function of $\mu_0$H and temperature $T$ at $\nu = -1$ and $\nu = -3$, respectively. The ferromagnetism critical temperature $T_c$ at $\nu = -1$ is 5 K while $T_c$ at $\nu = -3$ is 4 K. {\bf c.} RMCD signal as a function of filling factor ($\nu$) and displacement field ($D/\epsilon_0$) at $T$ = 1.7 K (left), $T$ = 2.6 K (mid), and $T$ = 3.5 K (right). The $\nu = -1$ state persists above $T$ = 3.5 K while the $\nu = -3$ state is suppressed.}
\label{fig:rmcdtemp}
\end{figure*}

\begin{figure*}[ht!]
\centering
\includegraphics{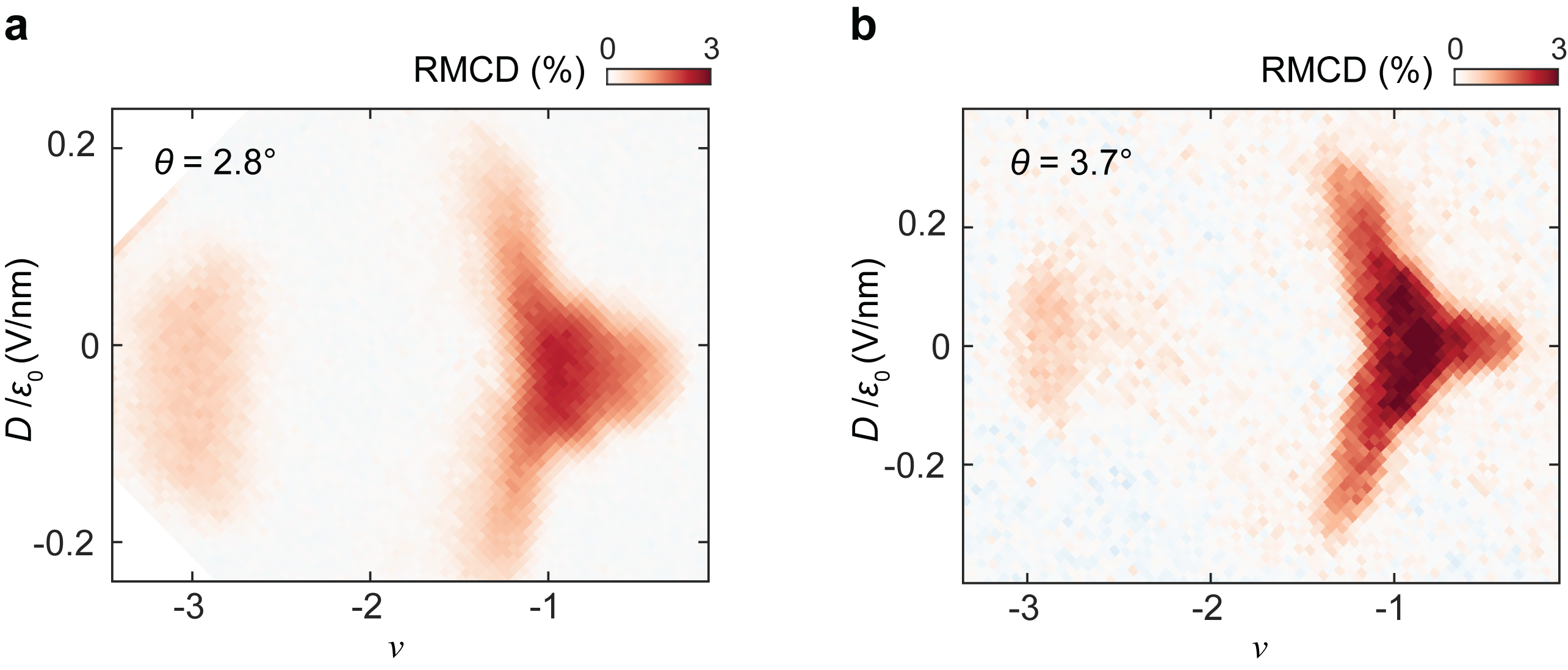}
\caption{\textbf{RMCD measurements of other twist angle tMoTe$_2$.} a, RMCD signal as a function of filling factor ($\nu$) and displacement field ($D/\epsilon_0$) for a, 2.8$\degree$, and b, 3.7$\degree$ samples, respectively.}
\label{fig:rmcdother}
\end{figure*}

\begin{figure*}[ht!]
\centering
\includegraphics{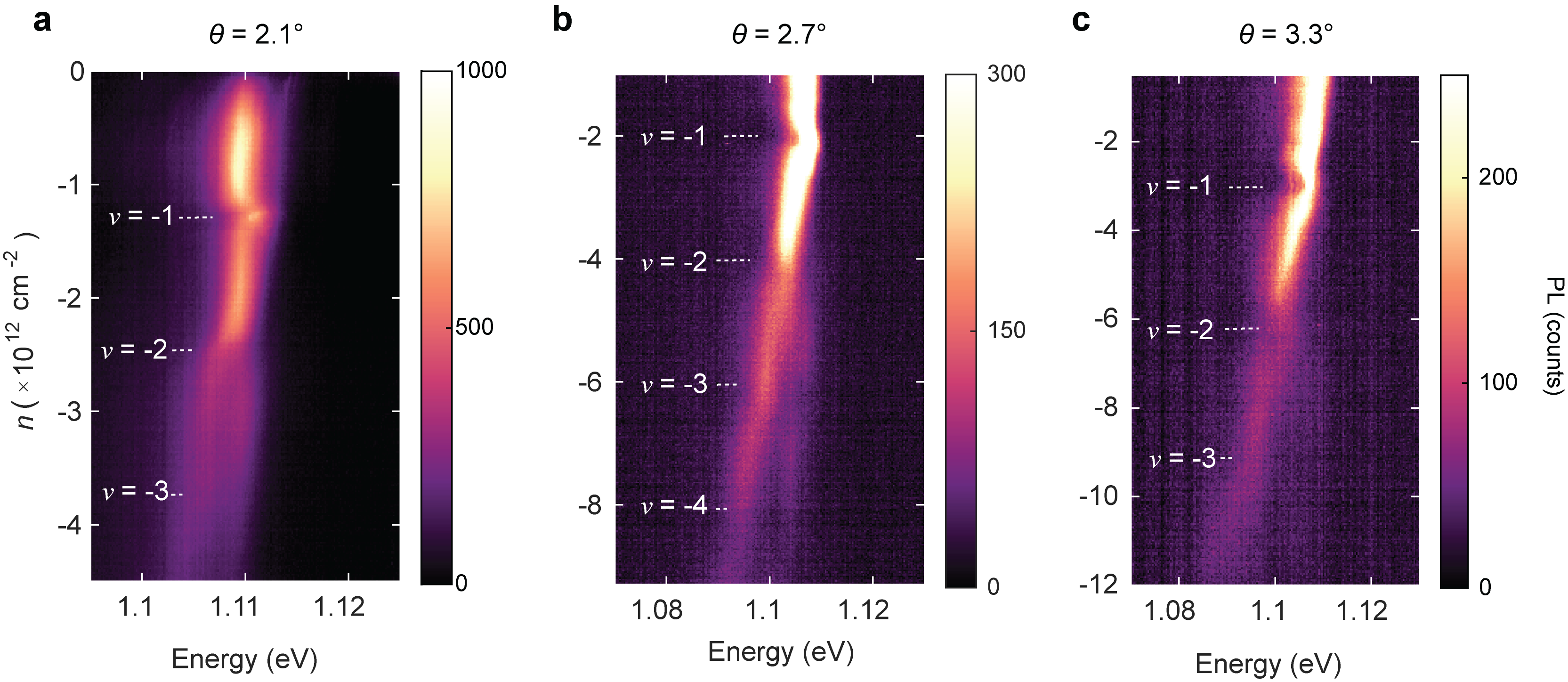}
\caption{\textbf{Doping dependent PL at other twist angle tMoTe$_2$}. Photoluminescence intensity plots as a function of photon energy and filling factor ($\nu$) for a, 2.1$\degree$, b, 2.7$\degree$, and c, 3.3$\degree$ tMoTe$_2$, respectively. A second, blue shifted peak can be observed emerging for hole doping above $\nu = -2$ in all three plots. }
\label{fig:PLother}
\end{figure*}

\begin{figure*}[ht!]
\centering
\includegraphics{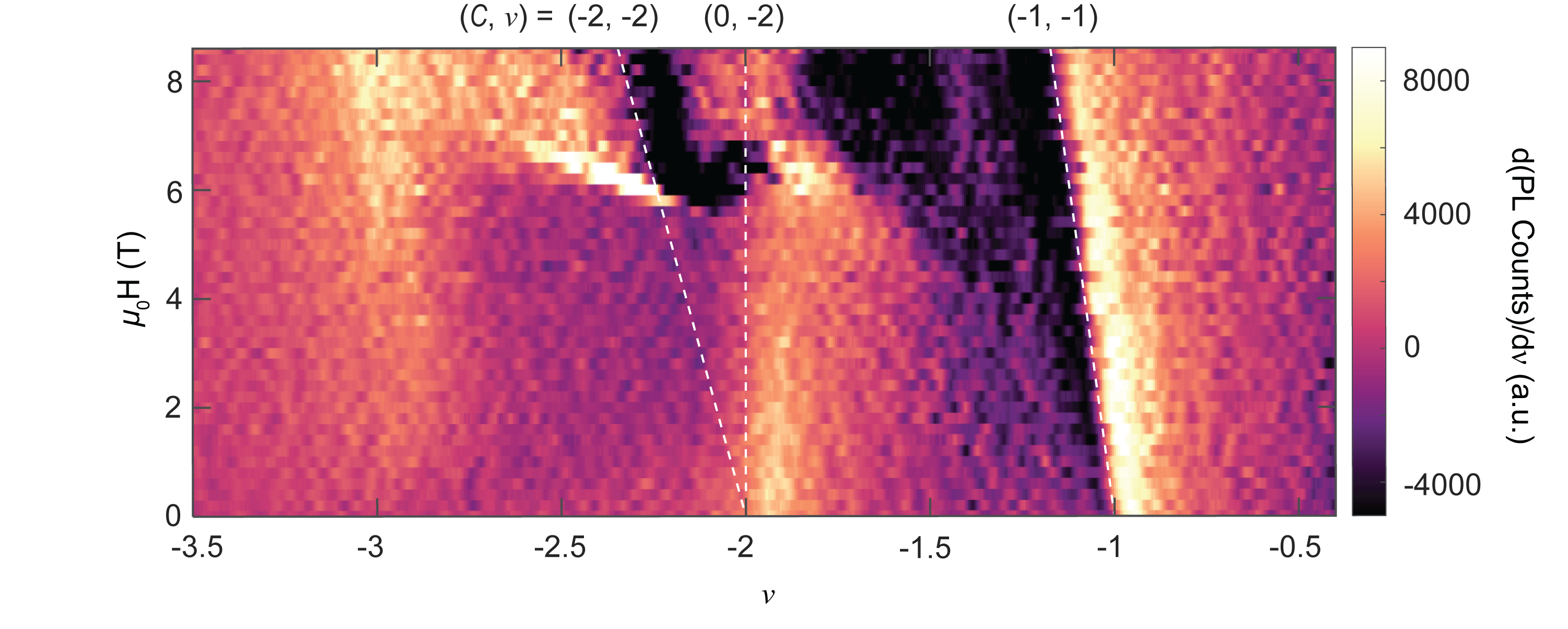}
\caption{\textbf{Derivative of integrated PL intensity versus $\nu$ and \textit{\textbf{$\mu_0$}H} for 2.1$\degree$ tMoTe$_2$.} Derivative of spectrally integrated PL intensity (dPL/d$\nu$) versus ($\mu_0$H) and filling factor ($\nu$) with 1 meV energy integration window. The dashed white lines are Wannier diagram corresponding to a $C$ = -1 QAH state at $\nu$ = -1 and a magnetic field induced $C = -2$ Chern insulator state at $\nu = -2$.}
\label{fig:opticalfanderivative}
\end{figure*}

\begin{figure*}[ht!]
\centering
\includegraphics{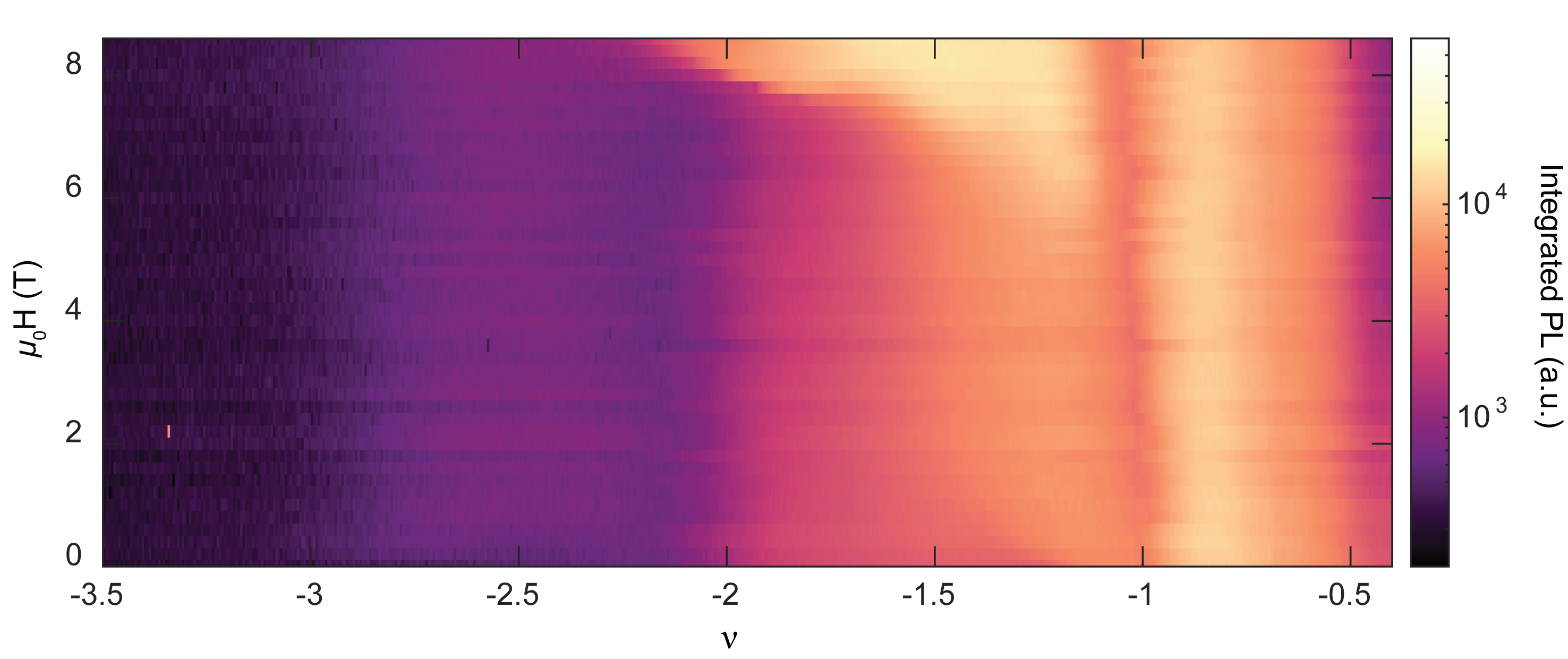}
\caption{\textbf{Optically detected fan diagram for  2.8$\degree$ tMoTe$_2$.}  Spectrally integrated PL intensity versus ($\mu_0$H) and filling factor ($\nu$) with 1 meV energy integration window in 2.8$\degree$ tMoTe$_2$.}
\label{fig:opticalfanother}
\end{figure*}

\end{document}